\documentclass[1p,twocolumn]{elsarticle}
 \pdfoutput=1
\usepackage{graphicx}
\usepackage{amsmath}
\usepackage{amssymb}
\usepackage{color}
\usepackage{verbatim}
\usepackage{url}
\usepackage{bm}
\usepackage{multirow}
\usepackage{booktabs}
\usepackage{rotating}
\usepackage{lipsum}
\usepackage{float}
\journal{.}

\begin{document}

\begin{frontmatter}

\title{Extraction of built-in potential of organic diodes from current-voltage characteristics}

\author{Prashanth Kumar M}
\author{Saranya R}
\author{Soumya Dutta}
\cortext[mycorrespondingauthor]{Corresponding author}
\ead{s.dutta@ee.iitm.ac.in}
\address[mymainaddress]{Department of Electrical Engineering, Indian Institute of Technology Madras, Chennai 600036, India}

\begin{abstract}
Physics based analytic equations for charge carrier profile and current density are derived by solving the carrier transport and the continuity equations for metal-intrinsic semiconductor-metal diodes. Using the analytic models a physics based method is developed to extract the built-in potential $V_{bi}$ from current density-voltage ($J$-$V$) characteristics. The proposed method is thoroughly validated using numerical simulation results. After verifying the applicability of the proposed theory on experimentally fabricated organic diodes, $V_{bi}$ is extracted using the present method showing a good agreement with the reported value.
\end{abstract}

\begin{keyword}
Organic diode \sep Organic solar cell \sep Device simulation \sep Built-in potential \sep Injection limited current \sep Space charge limited current
%\MSC[2010] 	65Z05\sep  99-00
\end{keyword}

\end{frontmatter}

%\linenumbers
\section{Introduction}

In organic diodes and solar cells, organic semiconductors are typically sandwiched between two dissimilar metal electrodes. The electric field and the charge profile under equilibrium are governed by the metal contacts of these devices. Moreover, the work function difference of the electrodes determines $V_{bi}$ \cite{Vbiorigin}, which is a crucial metric for both organic diode and solar cells as far as device performance is concerned. However, the charges that get injected from the contacts cause an excess potential drop near the metal-semiconductor junctions \cite{Simmons}. This excess drop in potential leads to a reduction of $V_{bi}$ \cite{TdependentVbi} to a lower value, say $V_{bi}'$, which is typically measured instead of $V_{bi}$ for a given temperature $T$. 

In this work we propose a physics-based analytical model for estimating $V_{bi}'$ from $J$-$V$ characteristics which takes the effect of injected charge into account. Our model is further extended to establish a method for extracting $V_{bi}$ by using $T$ dependency of $V_{bi}'$. Finally, the method is employed to estimate $V_{bi}$ of poly(3-hexylthiophene) (P3HT):phenyl-C61-butyric acid methyl ester (PCBM) based organic diodes, which are fabricated in our laboratory. 
\section{Simulation and analytical results}
\begin{figure}[!ht]
	\centering
	\includegraphics[scale = 1.2]{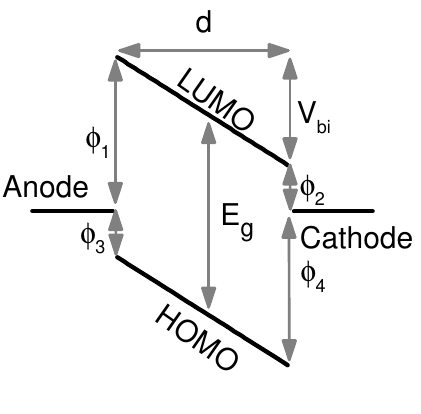}
	\caption{A schematic of equilibrium energy-band diagram of an organic diode, with organic semiconductor of thickness $d$ and bandgap $E_g$, where LUMO and HOMO being the lowest unoccupied and highest occupied molecular orbitals, respectively.}\label{Fig1Banddiagram}
	%	\vspace{-1.6em}
\end{figure}
We use the Metal-Insulator-Metal methodology for numerical simulation \cite{MIMBlom1,MIMBlom2,Koster}. The numerical simulations are done using the commercially available the Sentaurus technology computer-aided design (TCAD) tool \cite{Sentaurus}. In numerical simulations, we consider Schottky contacts at anode and cathode with barriers $\phi_1$ ($\phi_3$), $\phi_2$ ($\phi_4$) for electrons (holes) respectively [Fig. \ref{Fig1Banddiagram}].

Therefore the carrier concentrations at the contacts are determined by thermionic emission process and are given as \begin{equation}
\begin{array}{l} 
n_0=N_C\exp\left(-\frac{\phi_1}{q V_t}\right),\quad \ n_d=N_C\exp\left(-\frac{\phi_2}{q V_t}\right),
\\
p_0=N_V\exp\left(-\frac{\phi_3}{q V_t}\right),\quad \ p_d=N_V\exp\left(-\frac{\phi_4}{q V_t}\right),	\label{Eq1n_therm}
\end{array}
\end{equation} where $V_t$ is the thermal voltage, $n_0$ $(p_0)$, $n_d$ $(p_d)$ are the electron (hole) concentration at anode and cathode respectively, $N_C$ ($N_V$) is the effective density of states for electrons (holes).
\subsection{Classification of diodes}
At equilibrium (applied voltage $V=0$ V ) the dissimilar metal work-functions are aligned leading to band bending [Fig. \ref{Fig1Banddiagram}] which sets up a built-in electric field inside the device. The strength of the built-in electric field depends on $V_{bi}$, d and the injected charge. Depending on the magnitude of the injected charge and its effect on the electric field, the diodes can be classified into two categories: (1) Low space charge (LSC) case. (2) High space charge (HSC) case. The injected charge can be modified by changing $N_C$ ($N_V$), barrier for electrons (holes) and the temperature. In this particular study we keep $N_C$ ($N_V$) unchanged and vary the barriers for electrons (holes) and the temperature to explain LSC and HSC cases. The parameters associated with the simulation for LSC and HSC cases are given in Fig. \ref{Fig2_BDEf_LSC_HSC}.

\subsubsection{Low Space Charge case}
\begin{figure}[ht]
	\centering
	\includegraphics[scale=0.55]{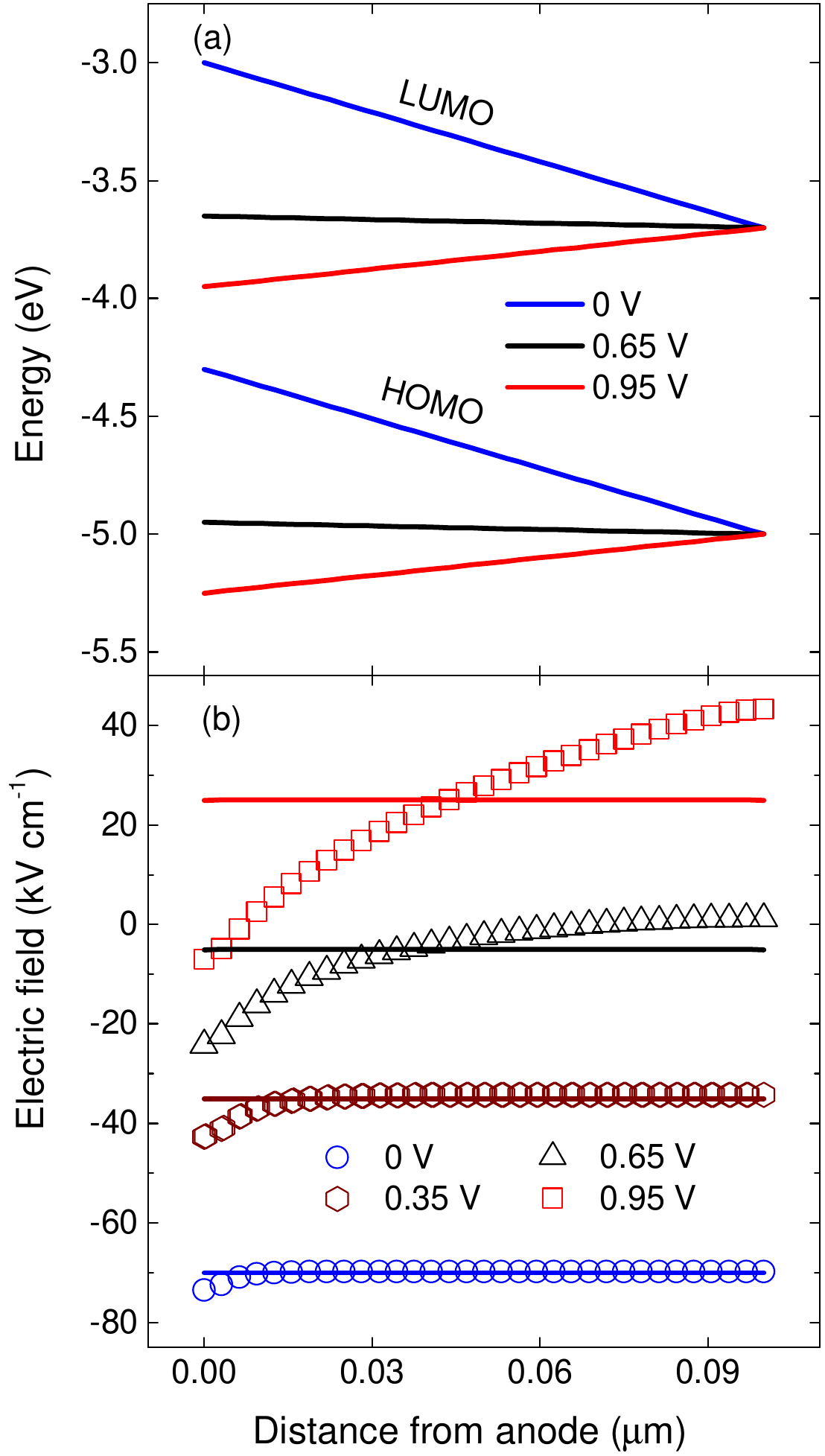}
	\caption{(a) The band diagram for LSC case for different $V$ (TCAD results), (b) Electric field profile of LSC (lines) and HSC (symbols) cases for different $V$. The parameters used for the simulation of LSC case are $E_g = 1.3$ eV, $\phi_1 = 1.0$ eV, $V_{bi} = 0.7$ V, $\mu_n = \mu_p = 1\times10^{-4}$ cm$^2$/Vs, $N_C = N_V=1\times10^{19}$ cm$^{-3}$, $\varepsilon$ = 3.3$\varepsilon_0$, $d=100$ nm and $T=300$ K. HSC case is resembled by changing $\phi_1$ to 1.15 eV and keeping all other parameters same as that of LSC.}\label{Fig2_BDEf_LSC_HSC}
	%	\vspace{-1.6em}
\end{figure}
In LSC case, the field due to the injected charge is very less compared to the electric field generated due to work-function difference. Hence the electric field is expected to be uniform by maintaining linear band bending [Fig. \ref{Fig2_BDEf_LSC_HSC}(b)] inside the device [Fig. \ref{Fig2_BDEf_LSC_HSC}(a)]. The injected carriers (from metals) undergo diffusion and drift concurrently in opposite direction to each other, as a consequence of concentration gradient and electric field respectively. Thus in order to model the carrier profiles and the current density, both drift and diffusion have to be considered simultaneously. The transport equation, describing electron current density, can be expressed as 
\begin{equation}
J_{n}= q n \mu_n E + q \mu_n V_t \frac{\partial n}{\partial x}, 
\label{Eq2Jna}
\end{equation} 
where $q$ is the electron charge, $E(x)$ is the electric field, $n(x)$ is the electron carrier concentration and $\mu_n$ is the mobility of electron. As discussed above, $E(x)$ is uniform and it is represented as \begin{equation}
E(x)  =   \frac{-\left(V_{bi}-V\right)}{d}.
\label{Eq3Efield}
\end{equation}
In order to arrive at analytic solution under steady state conditions, we consider three assumptions: (1) Semiconductor is intrinsic, (2) Carrier mobilities ($\mu_n$ and hole mobility $\mu_p$ ) are constant with respect to $V$ and $T$, (3) There is no carrier generation and recombination. The last assumption modifies the continuity equation for electrons as
\begin{equation}
\dfrac{\partial J_n(x)}{\partial x} = 0.
\label{Eq4nCont}
\end{equation} 	
Using Eqs. (\ref{Eq2Jna}), (\ref{Eq3Efield}) and (\ref{Eq4nCont}), a second order differential equation is developed for $n(x)$ as 
\begin{equation}
\dfrac{\partial^2 n(x)}{\partial x^2} + \dfrac{E(x)}{V_t}\dfrac{\partial n(x)}{\partial x} = 0. 
\label{cpldeq}
\end{equation}
By employing the thermionic emission boundary condition for electrons [Eq. \ref{Eq1n_therm}], an analytic solution for $n(x)$ is obtained as 
\begin{equation}
{\footnotesize n(x)=\frac{n_d-n_0\exp\left(\frac{V_{bi}-V}{V_t}\right)+\left(n_0-n_d\right)\exp\left(\frac{V_{bi}-V}{V_t}\frac{x}{d}\right)}{1-\exp\left(\frac{V_{bi}-V}{V_t}\right)}.
	\label{Eq6nx}}
\end{equation}
Similar approach can be used to obtain an analytic solution for holes.

\begin{figure}[ht] 	
	\centering
	\includegraphics[scale=0.55]{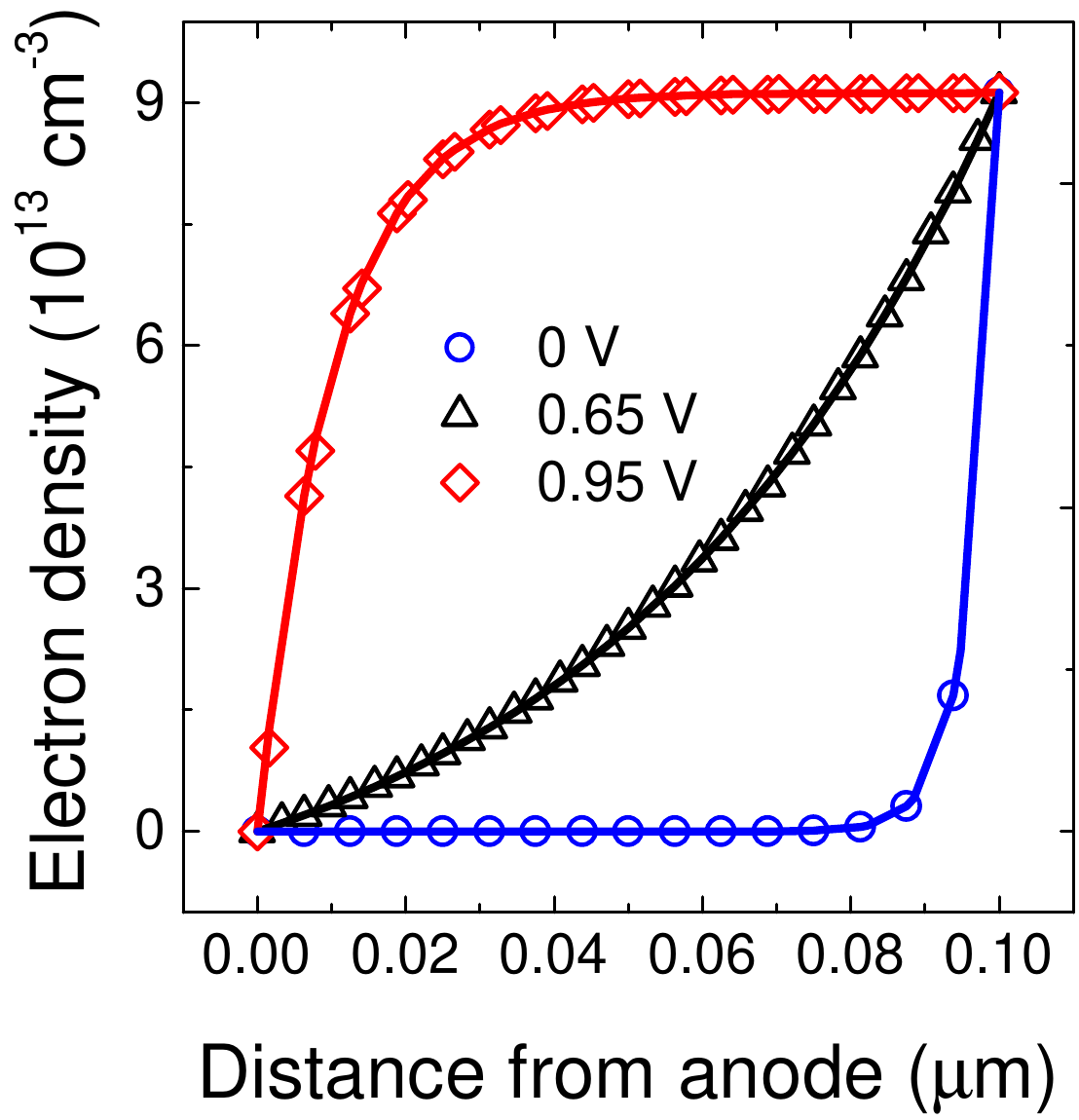}
	\caption{Electron profile for different $V$ as annotated, using TCAD simulation (symbols) and  Eq. \ref{Eq6nx} (solid lines).} 
	\label{Fig3_eprof}
	%	\vspace{-1.6em}
\end{figure}
The extracted electron profile inside the device using TCAD simulation (symbols) and Eq. \ref{Eq6nx} (solid lines) for different $V$ are compared in Fig. \ref{Fig3_eprof} which ensures that Eq. \ref{Eq6nx} is in good agreement with the TCAD results. Further, the analytic solution for current density can be expressed using Eqs. (\ref{Eq2Jna}), (\ref{Eq3Efield}), (\ref{Eq6nx}) and their hole counterparts as 
\begin{equation}
\begin{array}{l} 
J = \dfrac{q \left(\mu_n n_0 + \mu_p p_d\right) \left(V_{bi}-V\right)\left[\exp\left(\frac{V}{V_t}\right)-1\right]}{d\left[1-\exp\left(-\dfrac{V_{bi}-V}{V_t}\right)\right]}.
\label{J_LJ}
\end{array}
\end{equation}
\begin{figure}[ht] 	
	\centering
	\includegraphics[scale=0.6]{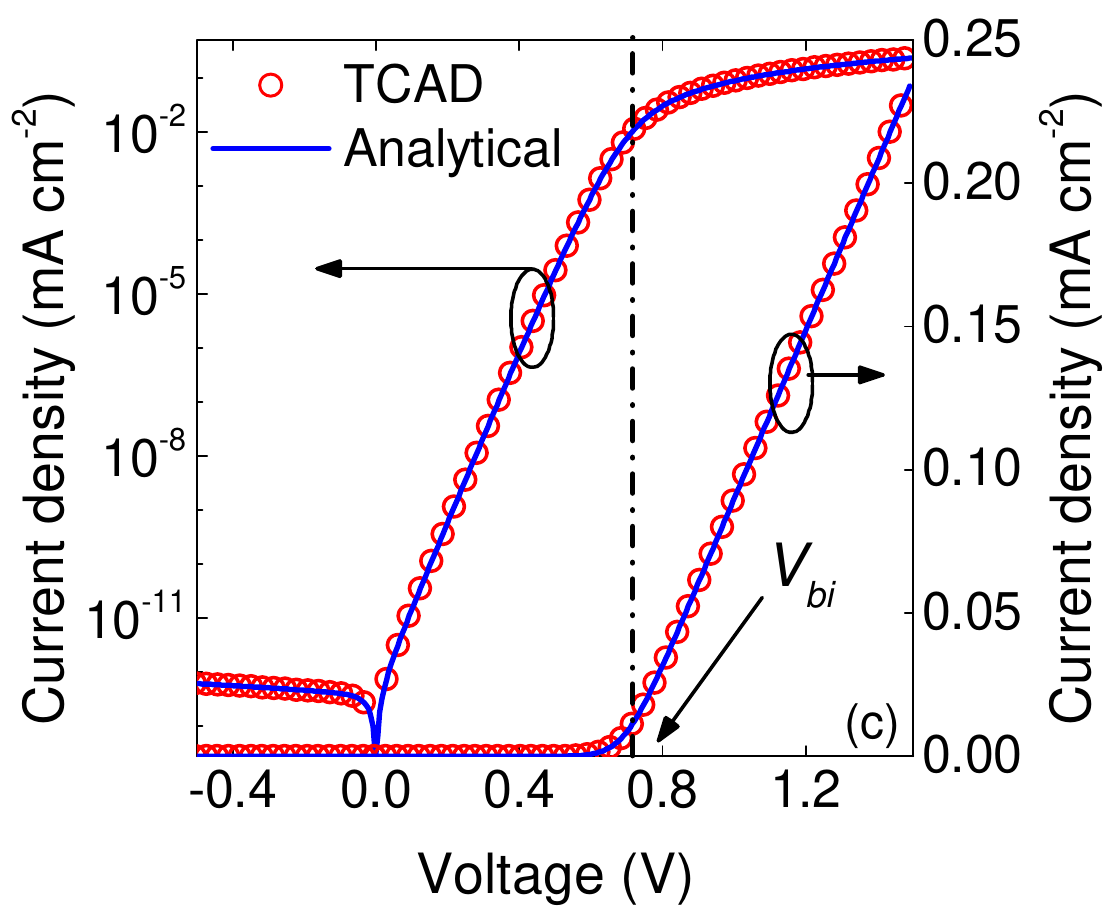}
	\caption{$J$-$V$ characteristics for LSC case where the symbols are TCAD and solid lines are model (Eq. \ref{J_LJ}).} 
	\label{Fig4JV_LSC}
	%	\vspace{-1.6em}
\end{figure}
The variation of $J$ with respect to $V$, based on TCAD simulation (symbols) and Eq. \ref{J_LJ} (solid line) are displayed in Fig. \ref{Fig4JV_LSC}, showing excellent consistency. A similar equation has been reported by different groups in the literature \cite{MIMBlom1,SJung}. Where S Jung et al. arrived at a similar analytical equation for less disordered organic materials with Gaussian density of states. However, the present method is completely rest upon charge based model with coherent device physics considering the effective density of states.

%\begin{figure*}[t]
%	\centering
%	\includegraphics[width=16.5 cm,trim = 30.0mm 25mm 1.07mm 35mm]{Fig3GV_LSC_HSC_Vbi_T1}
%	\caption{(a), (b) $G$-$V$ characteristics for LSC and HSC cases, (c) $V_{bi}'$ variation with respect to $T$ (inset shows the $V_\alpha$ variation) for different $V_{bi}$, varying from 0.6 V to 0.85 V insteps of 0.05 V with $\phi_2 = 0.45$ eV, (d) variation of extracted $V_{bi}$ and $\phi_3$ with respect to $\phi_1$, for different $\phi_2$ (comparison between TCAD results (symbols) and model (solid lines)).} 
%	\label{Fig3GV_LSC_HSC_Vbi_T}
%	%	\vspace{-1.6em}
%\end{figure*}
According to Fig. \ref{Fig3_eprof}, charge increases exponentially from anode to cathode for $V < V_{bi}$ (0.7 V in particular). The exponential nature of charge along with linear variation of $E(x)$ with $V$ results in exponential variation of $J$ with respect to $V$. On the other hand, for $V > V_{bi}$, the charge carrier profile changes significantly [Fig. \ref{Fig3_eprof}] by virtue of field reversal [Fig. \ref{Fig2_BDEf_LSC_HSC}]. Charge carrier concentration increases from anode to cathode like a logistic function [Fig. \ref{Fig3_eprof}], maintaining its spatially uniform nature inside the device except near the anode-semiconductor junction. The uniform nature of both charge carrier concentration and electric field profiles leads to a linear variation of current. In LSC case, the current is typically injection limited current (ILC) since dominant part of the current is controlled by the injected charge carriers.

\subsubsection{High Space Charge case}
\begin{figure}[t] 	
	\centering
	\includegraphics[scale=0.6]{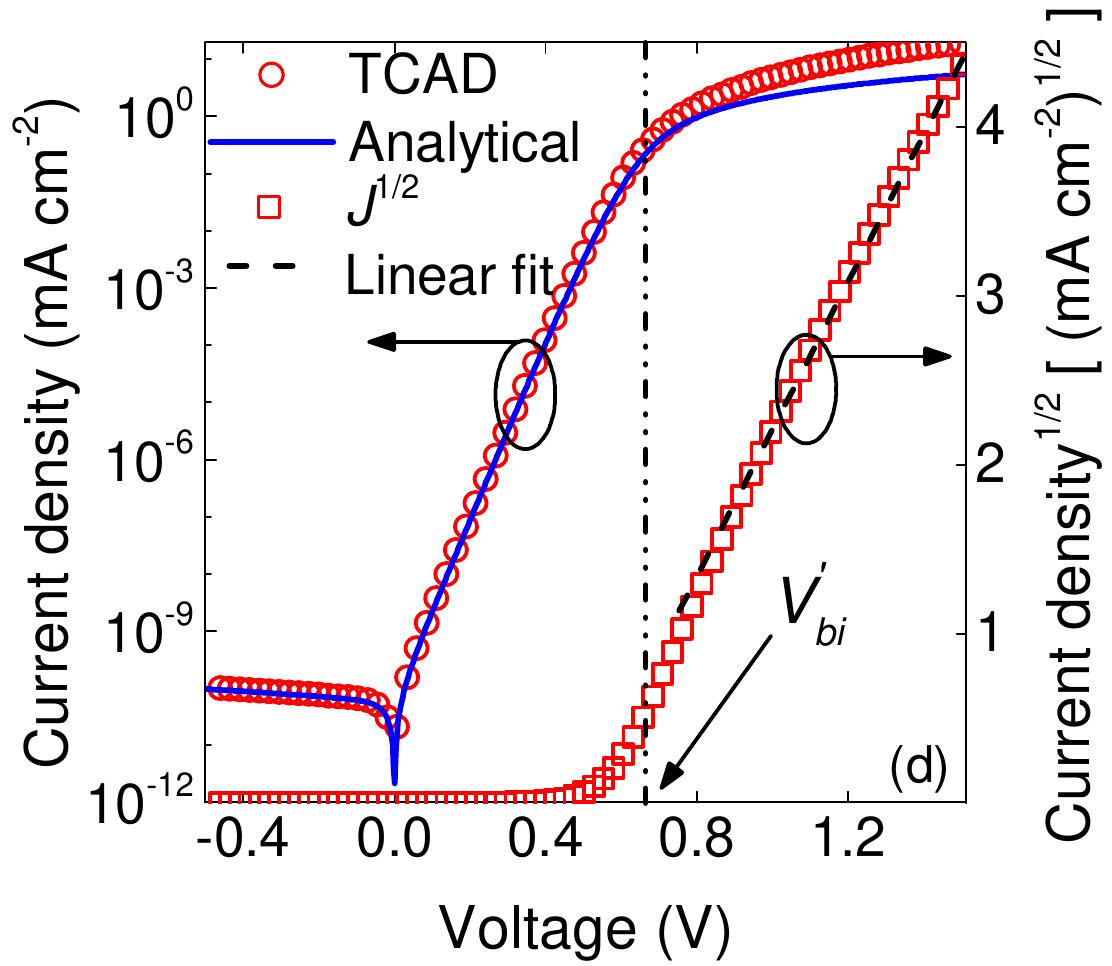}
	\caption{$J$-$V$ characteristics for HSC case, where the symbols are TCAD and solid lines are model (Eq. \ref{Eq9J_HJ}).} 
	\label{Fig5JV_HSC}
	%	\vspace{-1.6em}
\end{figure}

In HSC case, the electric field due to the injected charge becomes comparable to the electric field associated with band bending, expressed by Eq. \ref{Eq3Efield}. Hence the net electric field within the device becomes non-uniform. HSC case can be realized by reducing the barrier height for electrons (holes) or by increasing $N_C$ ($N_V$) or by increasing the thickness of the semiconductor. However, in this study HSC is realized by reducing the barrier height at anode-semiconductor junction in particular. Under equilibrium, a uniform electric field is observed within the device except near the anode-semiconductor junction where injected charge is high [Fig. \ref{Fig2_BDEf_LSC_HSC}(b)]. However, for $V < V_{bi}$ the magnitude of uniform electric field is slightly less than that of LSC case. Thus, $J$-$V$ characteristics maintain the same exponential nature as that of LSC case, exhibiting a reduction in built-in potential [Fig. \ref{Fig5JV_HSC}]. Hence it is essential to modify the electric field in case of HSC by reducing $V_{bi}$ to $V_{bi}-\phi$ (i.e., $V_{bi}'$) where $\phi$ accounts for the reduction in $E(x)$ due to the injected charge. However, in case of $V > V_{bi}'$, the non-linearity in electric field profile near the anode-semiconductor junction becomes predominant upon applying voltage and spreads throughout the device differing drastically from that of LSC case. As a consequence, the electric field and the carrier concentration become interdependent, leading to non-linear $J$-$V$ characteristics. The current density varies with square of $V$ (for $V > V_{bi}'$) as evidenced by a linear nature of $\sqrt{J}$-V characteristics [Fig. \ref{Fig5JV_HSC}]. As the current is controlled by the space charge, it is space charge limited current (SCLC). 

For $V < V_{bi}'$, in the uniform electric field region, the electric field strength can be modeled as 
\begin{equation} 
E(x)  =   \frac{-\left(V_{bi}' - V\right)}{d}.
\label{Eq8Efield_Hj}
\end{equation}  
Using Eq. \ref{Eq8Efield_Hj}, Eq. \ref{J_LJ} can be modified as 
\begin{equation}
\begin{array}{l} 
J = \dfrac{q \left(\mu_n n_0 + \mu_p p_d\right) \left(V_{bi}'-V\right)\left[\exp\left(\dfrac{V}{ V_t}\right)-1\right]}{ d\left[1-\exp\left(-\dfrac{V_{bi}'-V}{ V_t}\right)\right]}.
\label{Eq9J_HJ}
\end{array}
\end{equation} 
$J$-$V$ characteristic using Eq. (\ref{Eq9J_HJ}) shows a good agreement with TCAD results under $V < V_{bi}'$ for $\phi$ = 0.0544 V [Fig. \ref{Fig5JV_HSC}(d)], where $\phi$ is obtained by fitting TCAD results with Eq. (\ref{Eq9J_HJ}).
\subsection{Extraction of Built-in potential}
In LSC case, $J$ changes its nature from exponential to linear for $V > V_{bi}$, whereas in HSC case, $J$ changes its nature from ILC to SCLC for $V > V_{bi}'$. Most of the practical organic diodes belong to HSC case. To understand more about current transition from exponential to linear or ILC to SCLC, we adopt a function $G$, proposed by Mantri et al. \cite{VbiESTRizvi}, where $G$ is defined as
\begin{equation}
G =\frac{\partial \ln(J)}{\partial \ln(V)}. 
\label{Eq10G}
\end{equation}
\begin{figure}[ht]
	\centering
	\includegraphics[scale  = 0.55]{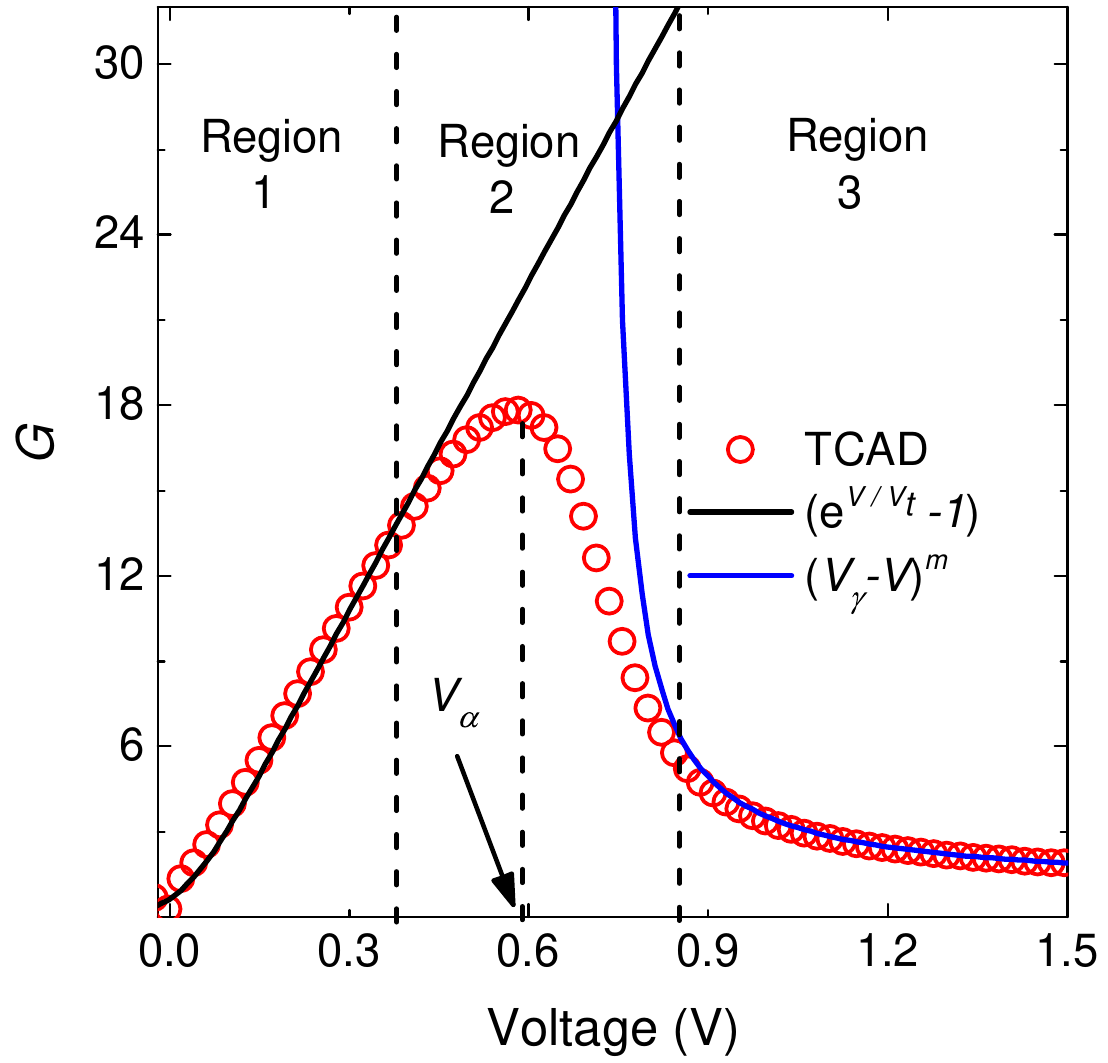}
	\caption{$G$-$V$ characteristics for LSC.} 
	\label{Fig6GV_schematic}
	%	\vspace{-1.6em}
\end{figure}
\begin{figure}[ht]
	\centering
	\includegraphics[scale  = 0.55]{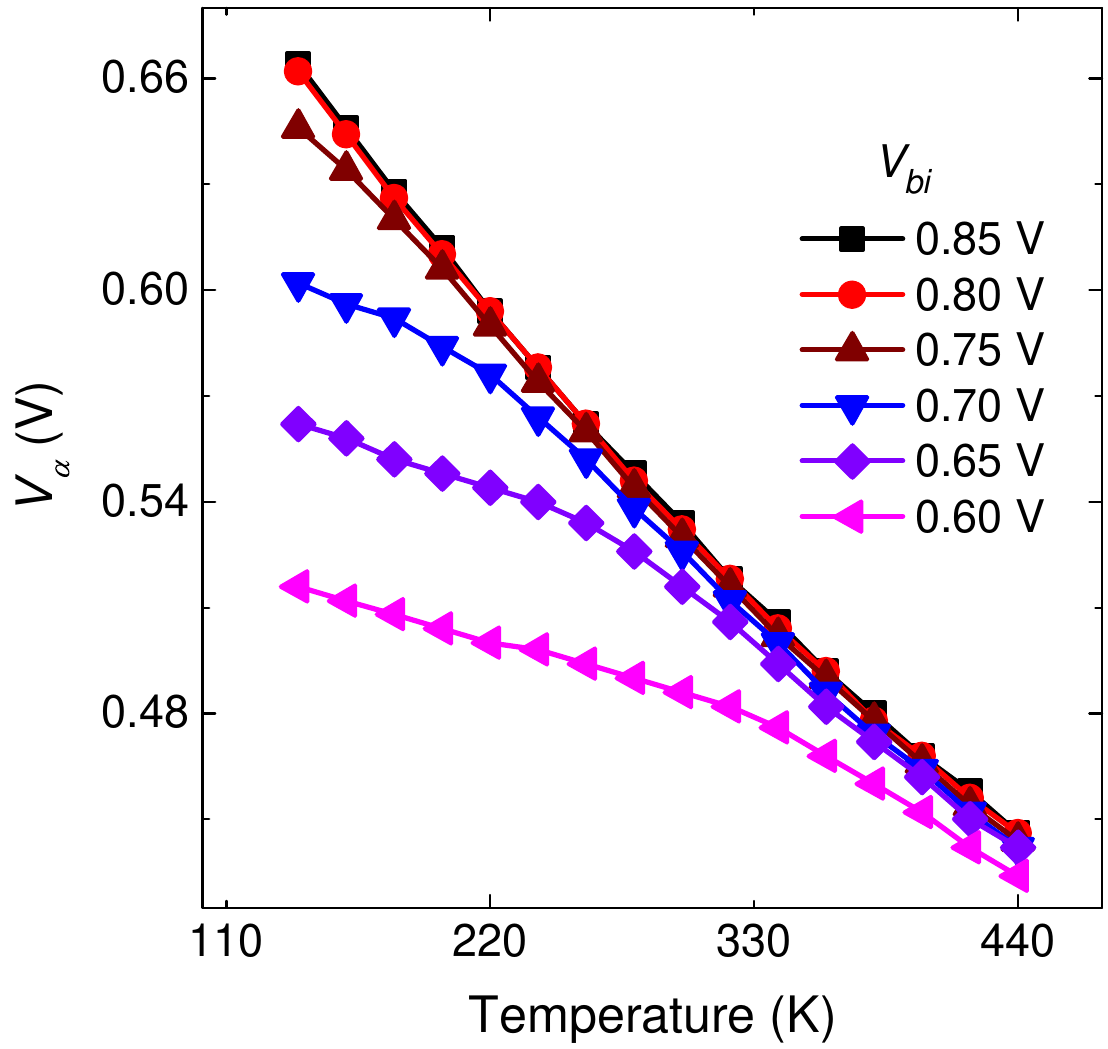}
	\caption{$V_\alpha$ variation with temperature for different $V_{bi}$ with $\phi_2 = 0.45$ eV (TCAD results).} 
	\label{Fig7_Vlpha_var_Temp}
	%	\vspace{-1.6em}
\end{figure}
The variation of $G$ with respect to $V$ shows three distinct regions signifying three different nature of current [Fig. \ref{Fig6GV_schematic}]. In region-1, the variation of $G$ can be fitted using a simple exponential function of $V$ as $[\exp(V/V_t)-1]$, as represented by solid line. In region-3, $G$ follows a power law as $(V_\gamma-V)^m$ with an exponent m, where $V_\gamma$ and $m$ are fitting parameters. Transition between these two regions (1 and 3) occurs through region-2, showing a combined effect of exponential and power law. Moreover, in region-2, $G$-$V$ characteristics exhibit a peak at a voltage, termed as $V_\alpha$. TCAD simulation results for the variation of $V_\alpha$ with respect to temperature as a function of different $V_{bi}$ is depicted in Fig. \ref{Fig7_Vlpha_var_Temp}. It emphasizes that modeling $V_\alpha$ variation with $V_{bi}$ will help in determining $V_{bi}$ from $J$-$V$ characteristics. Using Eq. (\ref{Eq9J_HJ}) and Eq. (\ref{Eq10G}), a unified expression for $G$ is developed as 
\begin{equation}
\begin{array}{l} 
G = \dfrac{V}{V_t \left[1-\exp\left(-\dfrac{V_{bi}'-V}{V_t}\right)\right]}+\dfrac{V}{V-V_{bi}'}.
\label{Eq11G_HJ}
\end{array}
\end{equation}
\begin{figure}[ht]
	\centering
	\includegraphics[scale  = 0.55]{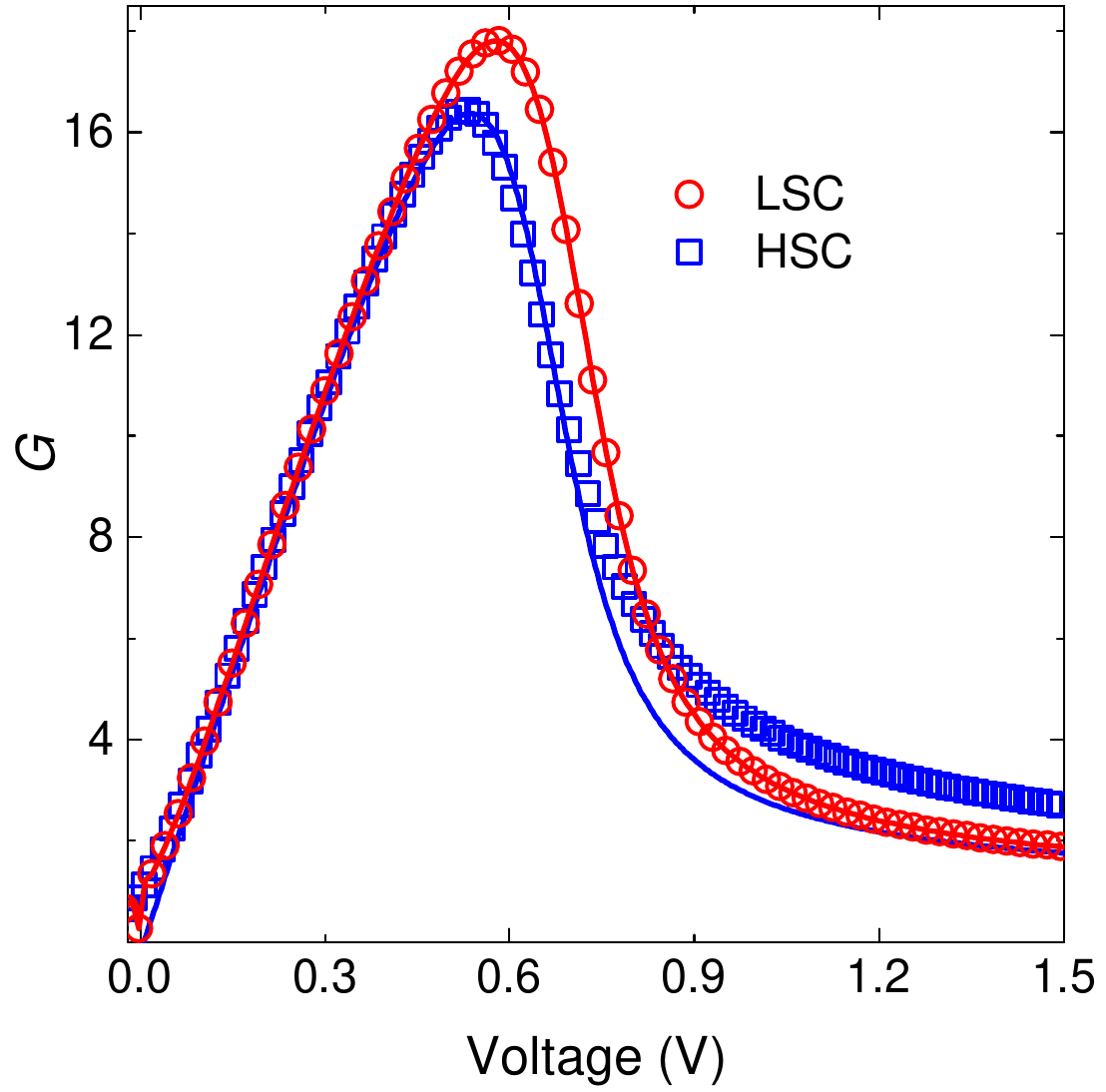}
	\caption{$G$-$V$ characteristics for LSC and HSC cases, where the symbols are TCAD and solid lines are model (Eq. \ref{Eq11G_HJ}).} 
	\label{Fig8_GV_LSC_HSC}
	%	\vspace{-1.6em}
\end{figure}
Eq. (\ref{Eq11G_HJ}) consists of two terms which correspond to two different nature of current. First term shows exponential nature, whereas the second term represents a power law with $V_\gamma = V_{bi}'$ and $m = 1$. Eq. (\ref{Eq11G_HJ}) shows an excellent match with the TCAD results for LSC case, where $V_{bi}'=V_{bi}$ [Fig. \ref{Fig8_GV_LSC_HSC}]. In case of HSC, there is indeed a good agreement between Eq. (\ref{Eq11G_HJ}) and TCAD results throughout regions 1 and 2. However, $G$-$V$ characteristics deviate from Eq. (\ref{Eq11G_HJ}) in region-3 due to the presence of SCLC, which cannot be captured by the present model. Thus Eq. \ref{Eq11G_HJ} is in good agreement with the TCAD results for both LSC and HSC cases throughout region-1 and region-2. Hence Eq. (\ref{Eq11G_HJ}) can be used for extracting $V_\alpha$ for both LSC and HSC cases. Using Eq. (\ref{Eq11G_HJ}) and equating its first derivative to zero at $V=V_\alpha$, we obtain 
\begin{equation}
\begin{array}{l} 
V_{bi}' V_t  - \left(V_{bi}'-V_\alpha \right)^2 \left[1 + \dfrac{V_\alpha}{V_t} \exp\left(-\dfrac{V_{bi}'-V_\alpha}{V_t}\right)  \right] = 0.
\label{Eq12dVG_eqn}
\end{array}
\end{equation} 
Eq. (\ref{Eq12dVG_eqn}) can be solved numerically to get $V_{bi}'$. For higher values of $V_{bi}$, the second term inside the square brackets of Eq. (\ref{Eq12dVG_eqn}) can be neglected. Therefore, a compact analytical equation is realized for $V_\alpha$ as
\begin{equation}
\begin{array}{l} 
V_\alpha = V_{bi}' - {\left(V_{bi}' V_t \right)}^{1/2}.
\label{Eq13VG_peak}
\end{array}
\end{equation} Subsequently $V_{bi}'$ is calculated as
\begin{equation}
\begin{array}{l} 
V_{bi}' =  \left(\dfrac{\sqrt{V_t}+\sqrt{ V_t+4 V_\alpha}}{2}\right)^2.
\label{Eq14Vbi_act}
\end{array}
\end{equation}
\begin{figure}[ht]
	\centering
	\includegraphics[scale  = 0.6]{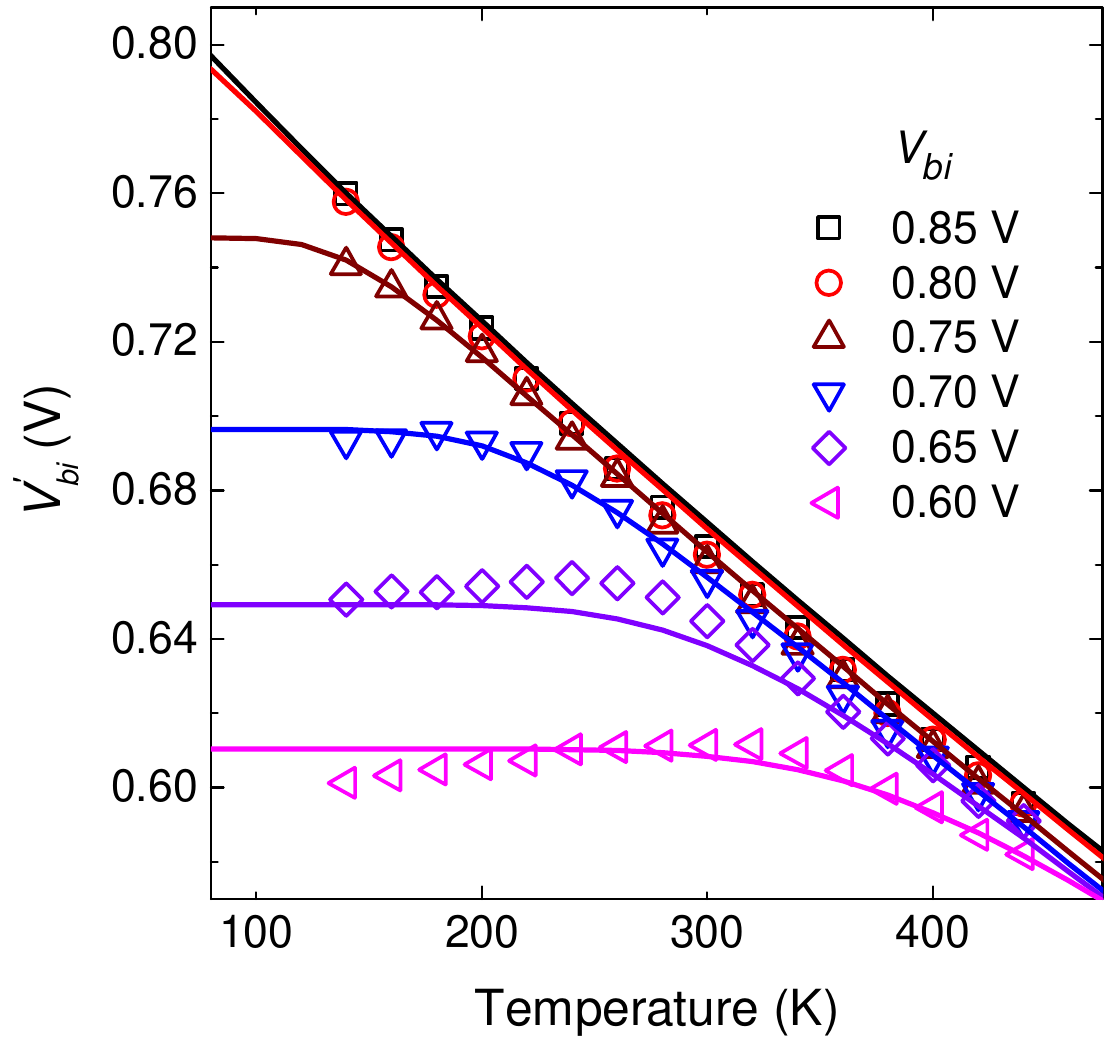}
	\caption{$V_{bi}'$ variation with respect to $T$ for different $V_{bi}$ with $\phi_2 = 0.45$ eV, where the symbols are TCAD and solid lines are model (Eq. \ref{Eq14Vbi_act}).} 
	\label{Fig9_Vbi_diff_temp_TCAD}
	%	\vspace{-1.6em}
\end{figure}
\begin{figure}[ht]
	\centering
	\includegraphics[scale  = 0.62]{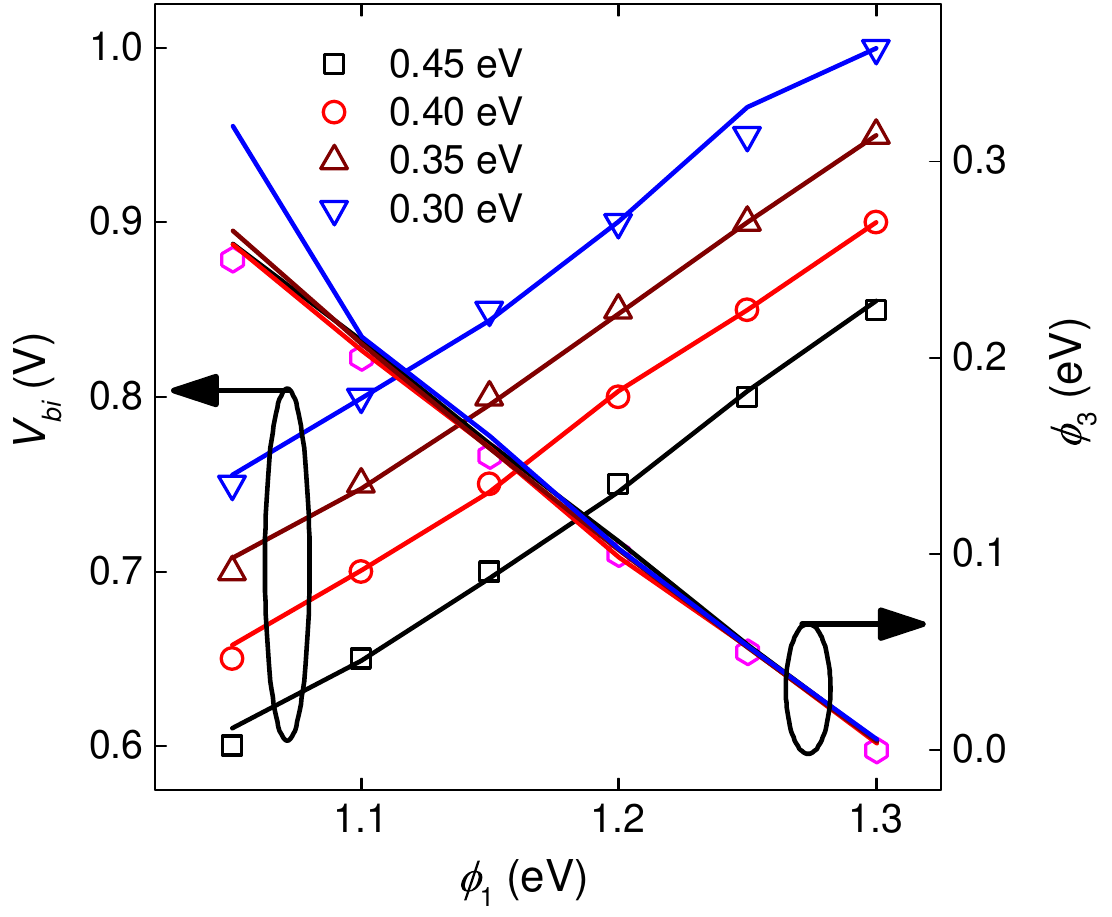}
	\caption{Variation of extracted $V_{bi}$ and $\phi_3$ with respect to $\phi_1$, for different $\phi_2$, where the symbols are TCAD and solid lines are model (Eq. \ref{Eq16Vbi'_act}).} 
	\label{Fig10_Vbi_phi3}
	%	\vspace{-1.6em}
\end{figure}
Using Eq. (\ref{Eq14Vbi_act}) and the value of $V_\alpha$ (extracted from $G$-$V$ characteristics), $V_{bi}'$ can be calculated. It is evident from Fig. \ref{Fig9_Vbi_diff_temp_TCAD} that $V_{bi}'$ increases with the decrease in $T$ and saturates to $V_{bi}$. The variation of $V_{bi}'$ with respect to $T$ arises due to $\phi$, which in turn depends on dominant injected charge near metal-semiconductor junction [$p_0=N_V\exp(-\phi_3/(k T))$ ] and thereby on $T$. Upon decreasing temperature, the amount of injected charge decreases, resulting $\phi$ tends to be zero and hence $V_{bi}'$ approaches to $V_{bi}$. $\phi$ can be calculated from the relation $\phi = V_{bi}-V_{bi}'$ for different $T$ and $V_{bi}$. From the variation of $\phi$ with respect to $T$, a semi-empirical model is developed for $\phi$ as

\begin{equation}
\begin{array}{l} 
\phi =k T \ln\left[\dfrac{q \varepsilon N_V\exp\left(-\frac{\phi_3}{k T}\right) }{r C_g^2 V_t}+1\right],
\label{Eq15phi}
\end{array}
\end{equation}
where $C_g=\varepsilon/d$, $\varepsilon$ is the dielectric constant and $r$ is a fitting parameter being independent of $T$. Using Eq. (\ref{Eq15phi}), $V_{bi}'$ can be written as \begin{equation}
\begin{array}{l} 
V_{bi}' = V_{bi} - k T \ln\left[\dfrac{q \varepsilon N_V\exp\left(-\frac{\phi_3}{k T}\right) }{r C_g^2 V_t}+1\right].
\label{Eq16Vbi'_act}
\end{array}
\end{equation}
$V_{bi}$ and $\phi_3$ can be obtained by solving Eq. (\ref{Eq16Vbi'_act}) self-consistently with $T$ dependent variation of $V_{bi}'$. The extracted parameters are in good agreement with TCAD results and it is validated for different combinations of $\phi_1$ and $\phi_2$, which shows the robustness of our model [Fig. \ref{Fig10_Vbi_phi3}].
%\subsection{Effect of Gaussian density of states}
%The above mentioned carrier concentration and current density analytical equations can be extended to density of states with Gaussian nature by modifying the $N_C$ and $N_V$ in Eq. \ref{Eq1n_therm}.    

\section{Experimental results}
\begin{figure}[htpb] 	
	\centering
	\includegraphics[scale =0.6 ]{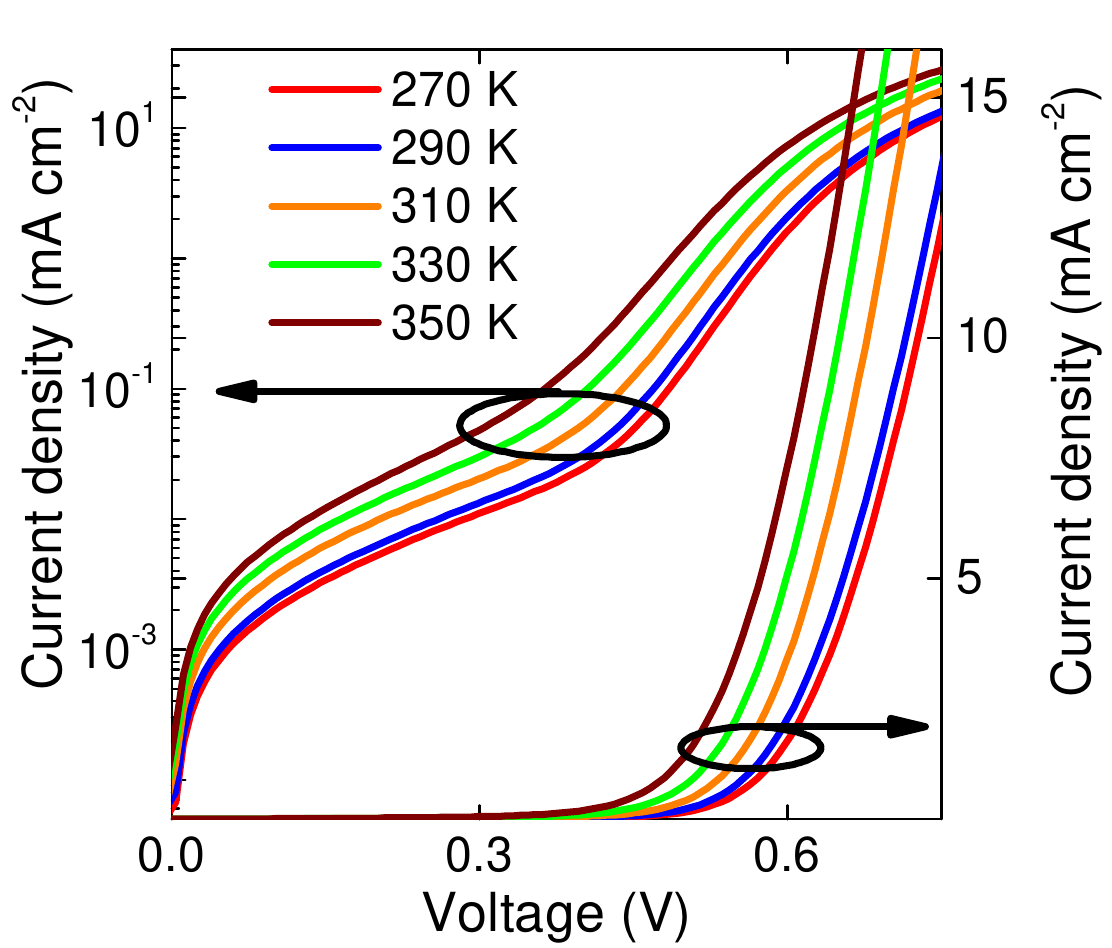}
	\caption{Experimental $J$-$V$ characteristics for device A.} 
	\label{Fig11_JV_exp}
	%	\vspace{-1.6em}
\end{figure}

In order to validate our model, $V_{bi}$ is extracted from experimental results of the organic solar cell, fabricated in our laboratory. Organic solar cells consisting of P3HT:PCBM as active material with aluminum (Al) as cathode and indium tin oxide (ITO)/poly(3,4-ethylene\-dioxythiophene):poly\-styrene sulfonate (PEDOT:PSS) as anode were fabricated inside a nitrogen glove box and characterized in a vacuum probe station. In order to validate the versatility of our model, we used three different thicknesses, 173 nm (Device A), 154 nm (Device B) and 106 nm (Device C) of P3HT:PCBM, resulting from the spin speed of 850 rpm, 1000 rpm and 1500 rpm respectively. $J$-$V$ characteristics of device A as a function of temperature is plotted in semilogarithmic scale and linear scale, as shown in Fig. \ref{Fig11_JV_exp}. It is observed that the forward current increases with increase in temperature as expected.
\begin{figure}[!h] 	
	\centering
	\includegraphics[scale =0.65]{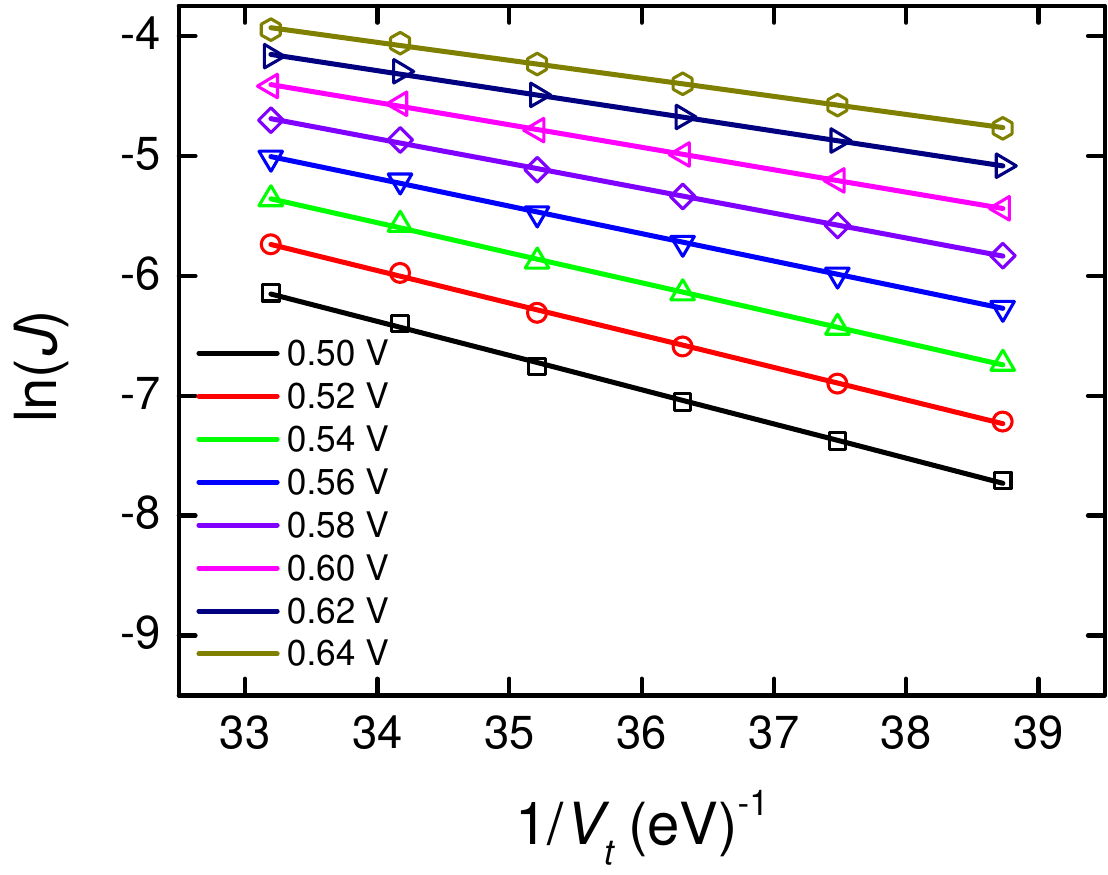}
	\caption{$\ln(J)$ variation with $1/V_t$ for different applied voltages, where symbols are experimental and solid lines are linear fit to the experimental data for device A.} 
	\label{Fig13_logJ_Vt}
	%	\vspace{-1.6em}
\end{figure}
\begin{figure}[H] 	
	\centering
	\includegraphics[scale =0.65]{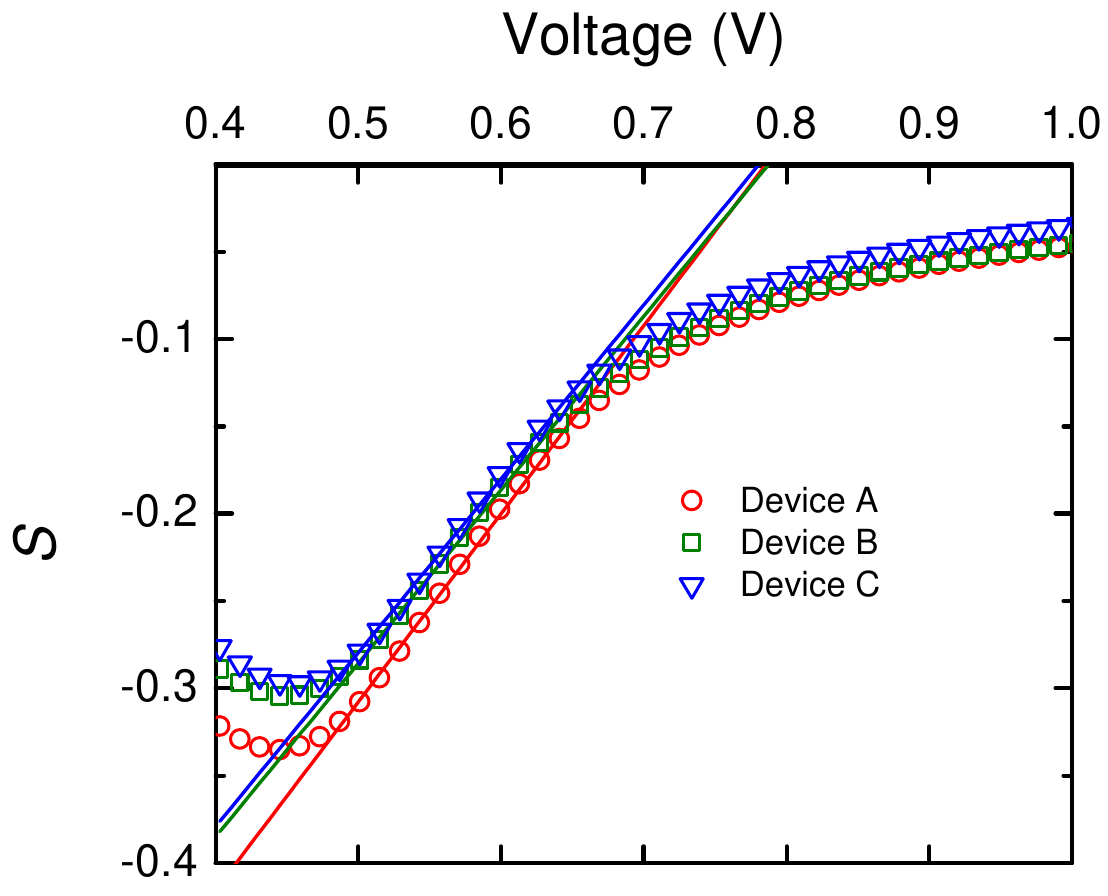}
	\caption{Variation of $S$ with applied voltage for different P3HT:PCBM spin speeds, where symbols are experimental and solid lines are linear fit to the experimental data. } 
	\label{Fig13_s_V}
	%	\vspace{-1.6em}
\end{figure}

\subsection{Model validation for experimental results}
To check whether the experimental results follow the proposed model, we have to analyze the $\ln\left(J\right)$ variation with $1/V_t$ for different $V$. For $0<V<V_{bi}'$, by considering $\mu_n n_0>\mu_p p_d$ and using Eq. \ref{Eq1n_therm} and Eq. \ref{Eq9J_HJ} $\ln\left(J\right)$ can be written as
\begin{equation}
\ln\left(J\right) = \ln\left[\dfrac{q \mu_n N_C \left(V_{bi}'-V\right)}{d}\right] + \dfrac{S}{ V_t},
\label{Eq9J_HJ_apprx}
\end{equation}
where
\begin{equation}
S=\frac{V-\phi_1}{\eta}.
\label{Eq18Svar}
\end{equation}
According to the proposed model $\ln\left(J\right)$ varies linearly with $1/V_t$ with a slope $S$ being dependent on $V$. According to Eq. (\ref{Eq18Svar}), $S$ varies linearly with $V$ having a slope $(1/\eta)$ equal to one and the intercept gives the value of one of the barrier potential ($\phi_1$). 
\begin{table}[H]
	\centering
	\caption{\label{table:1} The extracted parameters for P3HT:PCBM diode with different thickness.} 
	\vspace{2mm}
	%	\begin{ruledtabular}
	
	\begin{tabular}{ccccc}
		\toprule
		Device & $\eta$  & $\phi_{1/4}$ & $V_{bi}$ &$\phi_{2/3}$ \\
		&  &(eV)  & (V) &(eV)\\
		\midrule
		A   &0.9903   & 0.8087  & 0.660   & 0.2888 \\ 
		B   &1.006   & 0.7826  & 0.661   & 0.2871 \\ 
		C   &1.004   & 0.7779   & 0.654   & 0.2795  \\
		\bottomrule
	\end{tabular} 
		%	\end{ruledtabular}
\end{table}

The experimental variation of $\ln(J)$ for Device A (symbols) is shown in Fig. \ref{Fig13_logJ_Vt} and it confirm the linear variation of $\ln(J)$ with $1/V_t$ for different applied voltages. Hence the experimental results are fitted with linear variation to find $S$ and the intercept and this study is extended for Devices B and C. The variation of $S$ with $V$ is shown in Fig. \ref{Fig13_s_V}, one can notice from the figure that for $0.54<V<0.65$, $S$ varies linearly with $V$. Moreover the linear fit in that particular regime gives a slope which is nearly equal to one (Table \ref{table:1}), which is in consistent with the proposed model. Hence this confirm the applicability of the proposed model on these experimental results. In addition we extract one of the barrier potential $\phi_1$ (or $\phi_4$) for different devices which is nearly equal to $0.8$ eV as tabulated in Table \ref{table:1}.
\begin{figure}[H] 	
	\centering
	\includegraphics[scale =0.6 ]{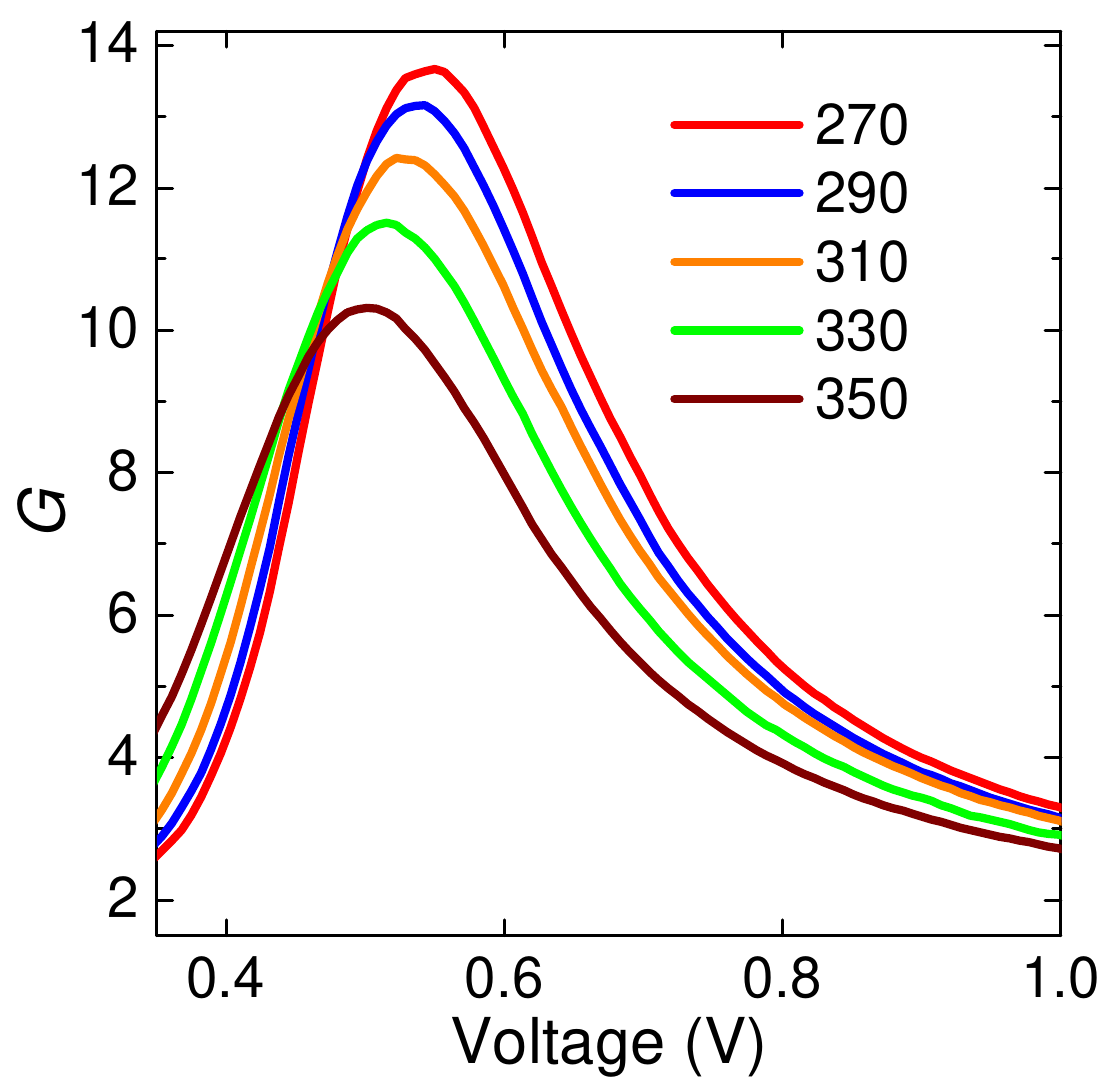}
	\caption{Experimental $G$-$V$ characteristics for device A.} 
	\label{Fig12_GV_exp}
	%	\vspace{-1.6em}
\end{figure}
\begin{figure}[H] 	
	\centering
	\includegraphics[scale =0.65 ]{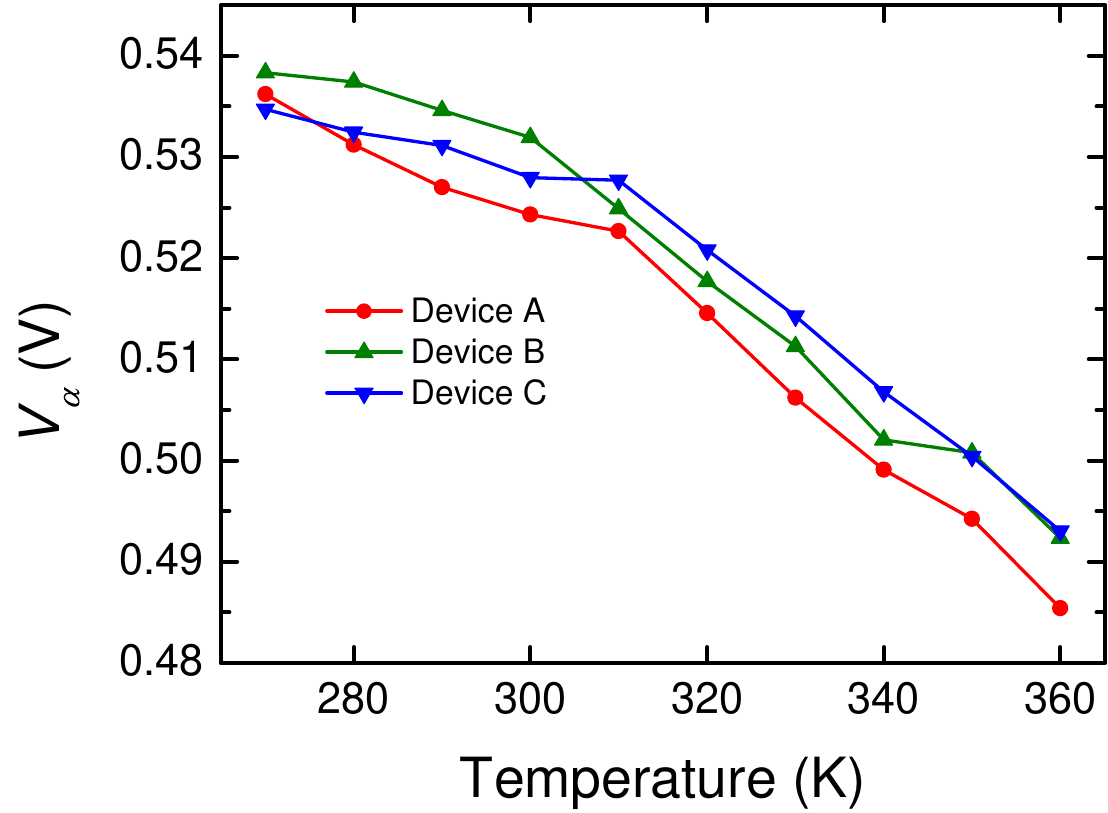}
	\caption{Experimental variation of $V_\alpha$ with temperature for devices with different P3HT:PCBM thickness.} 
	\label{Fig15_Valpha_exp}
	%	\vspace{-1.6em}
\end{figure}
Using Eq. (\ref{Eq10G}), we obtain $G$-$V$ plot for Device A, as shown in Fig. \ref{Fig12_GV_exp}. $V_\alpha$ is extracted from the peak position of $G$-$V$ plot for different temperatures and shown in Fig. \ref{Fig15_Valpha_exp} for different devices. Subsequently, $V_{bi}'$, calculated using Eq. (\ref{Eq12dVG_eqn}) for devices A,B and C, are shown in Fig. \ref{Fig13_Valpha_Vbi_exp} with symbols. As explained earlier, by solving Eq. (\ref{Eq16Vbi'_act}) in coherent manner, $V_{bi}$ and $\phi_3$ are extracted and presented in the Table \ref{table:1}.

\begin{figure}[H] 	
	\centering
	\includegraphics[scale =0.6]{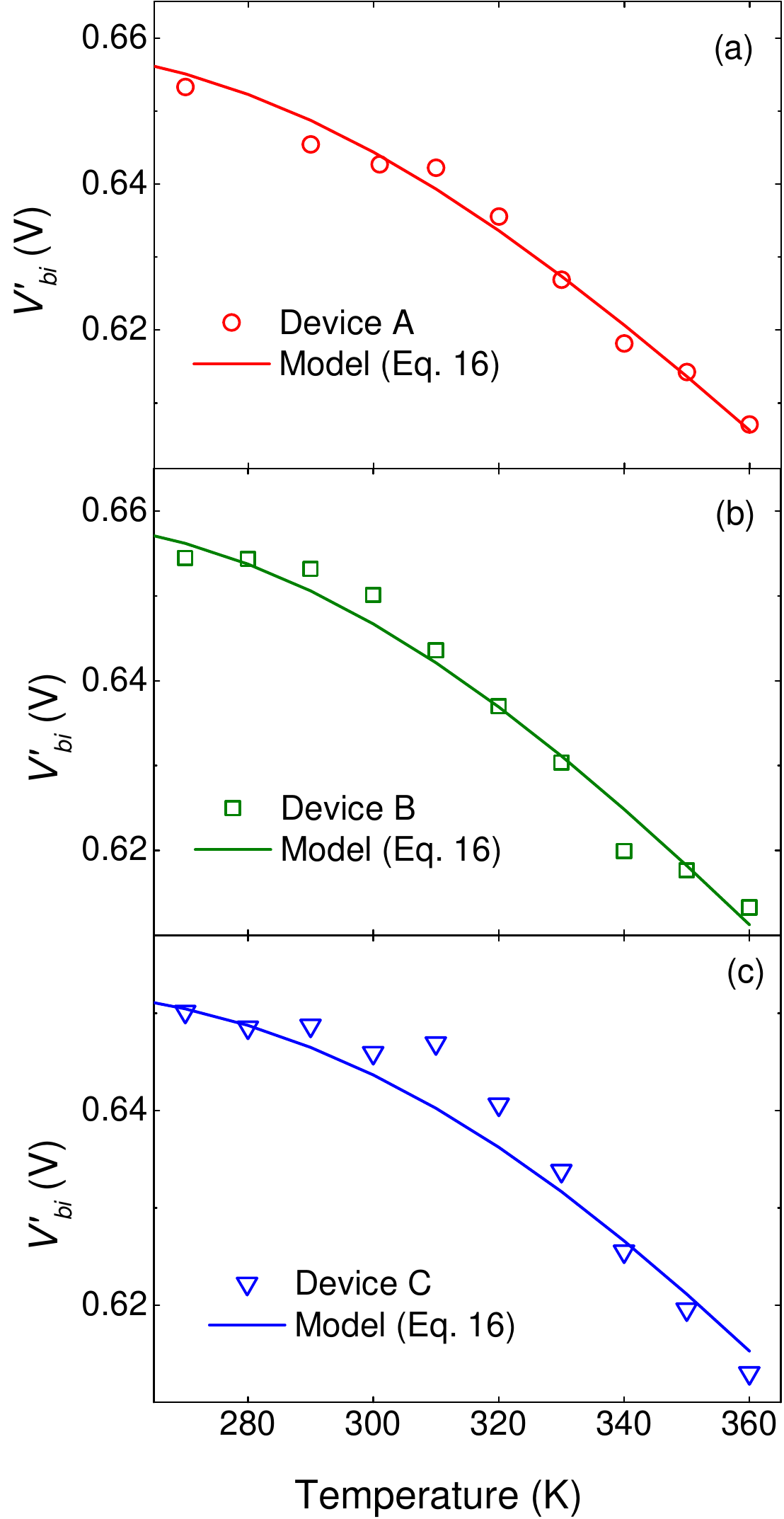}
	\caption{Experimental variation of $V_{bi}'$ with temperature for devices with different P3HT:PCBM thickness, where symbols are experimental and the solid lines are model.} 
	\label{Fig13_Valpha_Vbi_exp}
	%	\vspace{-1.6em}
\end{figure}

It is important to note that the extracted values of $V_{bi}$ are almost same for different thickness. Hence the $V_{bi}$ obtained using the present model is independent of thickness, as expected. Moreover the extracted values of $V_{bi}$ are in consistent with reported values for P3HT:PCBM device \cite{VbivalidityofMS}, which validates our model and ensures the method of extracting built-in potential from $J$-$V$ characteristics of organic diode or solar cell.
\section{Conclusion}
In summary, we developed analytic models for injected charge profile, $J$-$V$ characteristics and $V_{bi}'$. $V_{bi}$ is estimated using temperature dependent variation of $V_{bi}'$. The extracted values of $V_{bi}$ are in good agreement with TCAD results. The extracted $V_{bi}$ from experimental characteristics of P3HT:PCBM based solar cells (organic diode) is in good agreement with the reported values in the literature. 
\section*{Acknowledgements}
The authors would like to acknowledge Department of Science \& Technology (DST, Govt. of India), Nissan and IIT Madras for financial support and Centre for NEMS and Nanophotonics (CNNP) for providing fabrication facility.

%\bibliographystyle{ieeetr}
%\section*{Refernces}

\end{document}